\documentclass[preprint,12pt]{elsarticle}

\usepackage{graphicx}
\usepackage{amssymb}

\usepackage[utf8]{inputenc}
\usepackage{geometry}
\usepackage{float}
\usepackage{sectsty}
\sectionfont{\centering}
\usepackage{textcomp}
\journal{ASRC Proceedings}

\begin{document}

\begin{frontmatter}


\title{\textbf{Near-Infrared atmospheric modelling of Jupiter\textquotesingle s South Equatorial Belt (SEB) observed with AAT/IRIS2
}}

\author{B.$\enspace$Karamiqucham$^{1,2}$,$\enspace$J.$\enspace$A.$\enspace$Bailey$^{1,2}$,$\enspace$L.$\enspace$Kedziora-Chudczer$^{3}$and$\enspace$D.$\enspace$V.$\enspace$Cotton$^{1,2}$ 
}

\address{$^1$School$\enspace$of$\enspace$Physics,$\enspace$University$\enspace$of$\enspace$New$\enspace$South$\enspace$Wales,$\enspace$Sydney,$\enspace$NSW$\enspace$2052,$\enspace$Australia}

\address{$^2$Australian$\enspace$Centre$\enspace$for$\enspace$Astrobiology,$\enspace$University$\enspace$of$\enspace$New$\enspace$South$\enspace$Wales,$\enspace$Sydney,$\enspace$NSW$\enspace$2052,$\enspace$Australia 
}

\address{$^3$Computational$\enspace$Engineering$\enspace$and$\enspace$Science$\enspace$Research$\enspace$Centre,$\enspace$University$\enspace$of$\enspace$Southern$\enspace$Queensland\\
$\enspace$Darling$\enspace$Heights,$\enspace$QLD$\enspace$2109,$\enspace$Australia}

\begin{abstract}
Near-Infrared spectra of Jupiter\textquotesingle s South Equatorial Belt (SEB) with AAT/IRIS2 in the H and K bands at a resolving power of R $\sim$ 2400 have been obtained. By creating line-by-line radiative transfer models with the latest improved spectral line data for ammonia and methane (HITRAN2016), we derive best models of cloud/haze parameters in Jupiter\textquotesingle s South Equatorial Belt. The modelled spectra fit the observations well except for small, isolated discrepancies in the trough region of H$_{2}$-H$_{2}$ collision-induced-absorption around 2.08 $\mu$m and the methane absorption level between 2.16 and 2.19 $\mu$m in K band and at the high pressure methane window between 1.596 to 1.618 $\mu$m in H band. 

\end{abstract}


\end{frontmatter}


\section{Introduction}
\label{S:1}

The near-Infrared spectra of Jupiter can be explained by different vertical structure of clouds and their varied composition. These spectral features are dominated by ammonia and methane absorption in Jupiter\textquotesingle s hydrogen rich atmosphere. Formation and structure of absorption bands can be interpreted as scattered/reflected Sun light from clouds at different atmospheric (tropospheric) pressure levels. 

Previously published models [e.g. 1, 2, 3, 4, 5, 6, 7, 8, 9] suggest three cloud decks in Jupiter\textquotesingle s upper atmosphere, which have different heights and opacities depending on their latitudinal location (zones, belts and polar). They are located at approximately 7 bar (water cloud), 2-2.5 bar (ammonium hydrosulfide cloud) and 0.6-0.7 bar (ammonia cloud) "respectively" from lower to higher altitudes in the atmosphere. In our models, we considered the upper two clouds with addition of a thin haze layer on top. 

\thispagestyle{empty}

Our main focus in this paper is the models of Jupiter\textquotesingle s near-Infrared H and K spectral bands. These spectral regions probe the absorbed/scattered sunlight of different particles, aerosols and gas molecules in the atmosphere, from the stratospheric haze down to the upper tropospheric clouds. Each band is sensitive to different particle sizes and cloud optical depths, therefore different atmospheric heights will influence the spectra. K band is mostly sensitive to the level of upper tropospheric/stratospheric haze and H band to the upper tropospheric pressures down to around 1 bar.

In this study we characterised cloud/haze variations in Jupiter\textquotesingle s SEB, focusing on the cloud pressure heights and optical depths. In our paper  we provide the observation and data reduction details in Section 2 followed by description of our VSTAR-ATMOF model settings in Section 3. The results and discussion explained and illustrated in Section 4 followed by a concluding section at the end.

\section{Observations and data reduction}
\label{S:2}

The data were obtained on 31st July 2010 with the InfraRed Imager and Spectrograph 2 (IRIS2; [10]) at the 3.9 m Anglo-Australian Telescope (AAT) (Table 1). The observations described here are long-slit (7.7 arcmin long and 1 arcsec wide) spectral 3-D cube of Jupiter at a relatively high resolving power of R $\sim$ 2400 in the H and K bands and spatial resolution of about 1405.2 km/pixel on the planet\textquotesingle s disk. 

The image of Jupiter and the selected spectra corresponding to the central meridian are shown in Figures 1 and 2 "respectively". In Figure 2, the dominant gaseous methane spectral window is present (e.g centred at ~1.600 micron) which make observing possible to the higher pressure (deeper layers) in the atmosphere. The areas are surrounded by regions of higher absorption and lower reflectance on their neighbouring wings. This is the case for the J and H bands deep enough to ignore the hydrogen collision-induced-absorption and only account for the gaseous opacity of methane. In K band, e.g. from ~2.04 micron to ~2.12 micron, (more obvious at the centre of the planet) the gaseous opacity of the atmosphere increases due to increasing absorption contributions from both methane and molecular hydrogen. 

\thispagestyle{empty}

For the purpose of this work we focused on the H and K bands from (1.47 to 1.82 $\mu$m) and (2.04 to 2.37 $\mu$m) "respectively" (Figure 3). Figure 4 shows different absorption levels of the atmospheric species in H and K band regions. Image (A) shows the CH$_{4}$ absorption band ($\sim$1.673 $\mu$m) along with the characteristic structure of reflective clouds in the atmosphere. The Southern Hemisphere is dimmer than the Northern, which is indicative of cloud decks at higher pressures (lower altitudes). Image (B) in Figure 4 illustrates the CH$_{4}$ absorption band ($\sim$2.189 $\mu$m) of mid-high atmospheric cloud/haze in both polar and equatorial regions. Image (C) in Figure 4 shows the spectral region affected by H$_{2}$-H$_{2}$ collision-induced-absorption around 2.12 $\mu$m which affects the planet disk with the exception of high stratospheric haze in both polar regions. 

The Figaro Package of Shortridge et al. [11] was used to process our data as described in Kedziora-Chudczer and Bailey [9]. 


\begin{table}[H]
\caption{Observations of Jupiter on 31 July 2010.
}
\centering
\resizebox{\textwidth}{!}{
\tabcolsep 1.4 pt
{\renewcommand{\arraystretch}{1.2} 
\begin{tabular}{|c|c|c|c|c|c|}
\hline
\hline
Spectral   &   Centre    & Mean Dispersion &   Start Time (UT)  & End Time (UT)  &  Total Exposure  \\
Bands   &   ($\mu$m)   &   (nm/pixel)  &   (h) &  (h)   &  (S)   \\
\hline
\hline
   
H  &   1.637   &   0.341   &   15:12   &   15:26    &   720 \\
\hline
K  &   2.249   &   0.442   &   14:37   &   14:52    &   720  \\
\hline
\hline

\end{tabular}
}
}

\label{tab:LR_results}
\end{table}


\begin{figure}[H]
\centering
\includegraphics[width=\columnwidth]{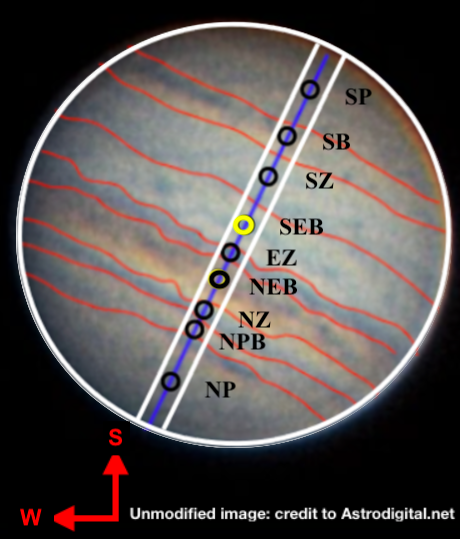}
\caption{Jupiter image from 31/07/2010, shows 9 latitudes separated by red lines provided to visualise the limits of zones and belts based on the visible image of the planet. The yellow circle shows the modelled South Equatorial Belt (SEB) in this study located at 16-6 degrees South planetographic latitude. Near-Infrared spectral models of the planet's whole disk (J, H and K bands) in all latitudes and longitudes will be the subject in our upcoming paper.
}
\end{figure}

\thispagestyle{empty}

\begin{figure}[H]
\centering
\includegraphics[width=\columnwidth]{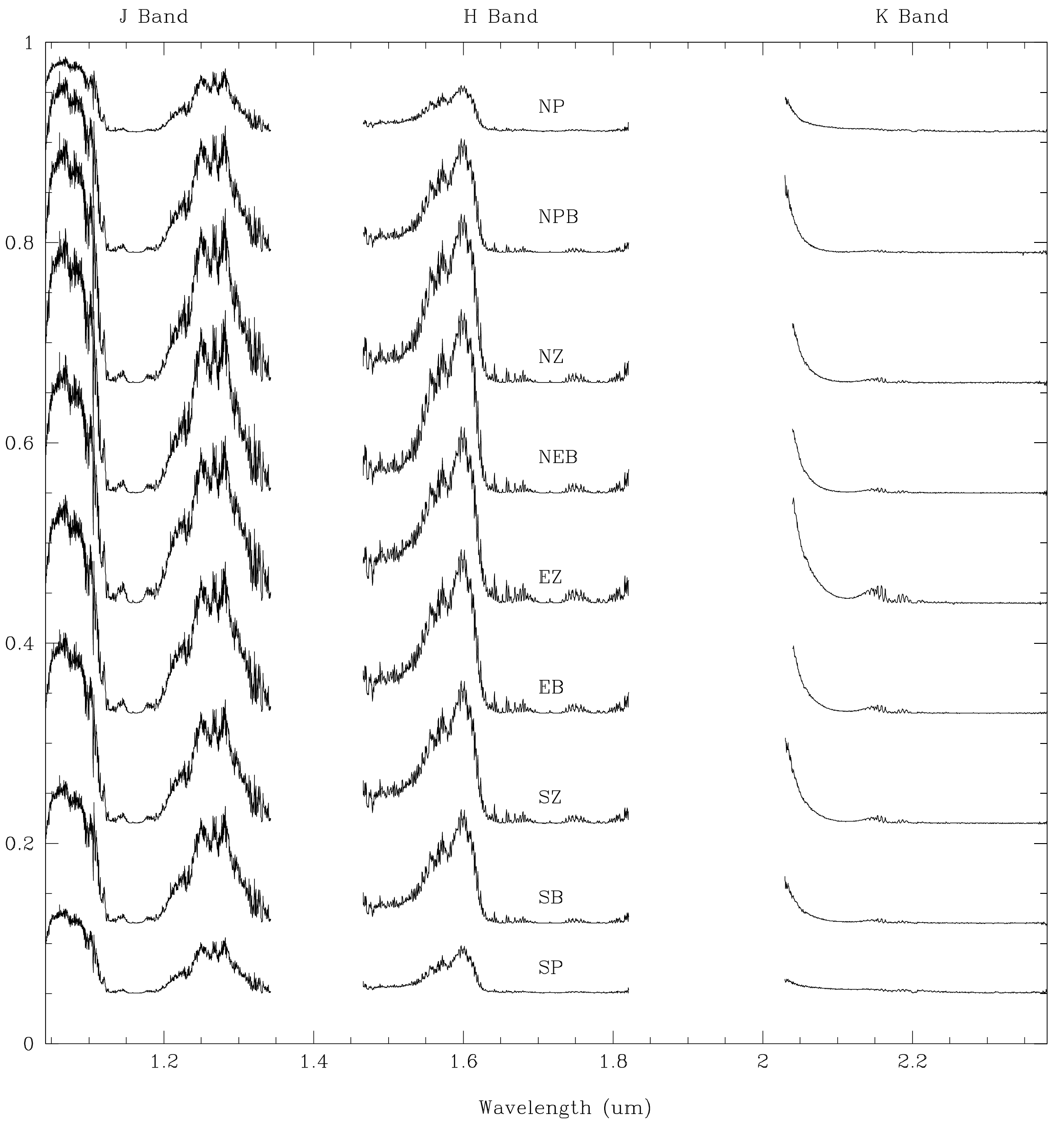}
\caption{The observed spectra of the planet in the J, H and K bands covering the area from 1.04 to 2.37 $\mu$m. The numbers presented in the vertical axis are for locating and aligning the spectra in the specified locations. The spectral region in H band ($\sim$1.47 to $\sim$1.82 $\mu$m) shows a gradual clearing in the amount of upper tropospheric haze (opacity) from the centre of the planet towards the polar regions. The thickest haze is visible in the equatorial zone and the thinnest in the poles. In the K band, the shape of the collisional-induced-absorption (centred around 2.08 $\mu$m) changes from the planet\textquotesingle s centre (EZ) towards the poles. This may suggest that the compressed (less vertically extended cloud) and optically thicker tropospheric cloud becomes thinner as we move poleward. The region centred around 2.15 $\mu$m is sensitive to the stratospheric haze and shows a poleward clearance of small sub-micron particles. The marked latitudes, from top to the bottom are: NP (north pole), NPB (northern polar belt), NZ (northern zone), NEB (northern equatorial belt), EZ (equatorial zone), SEB (EB) (south equatorial belt), SZ (southern zone), SB (southern belt) and SP (south pole). 
}
\end{figure}

\thispagestyle{empty}

\begin{figure}[H]
\centering
\includegraphics[width=\columnwidth]{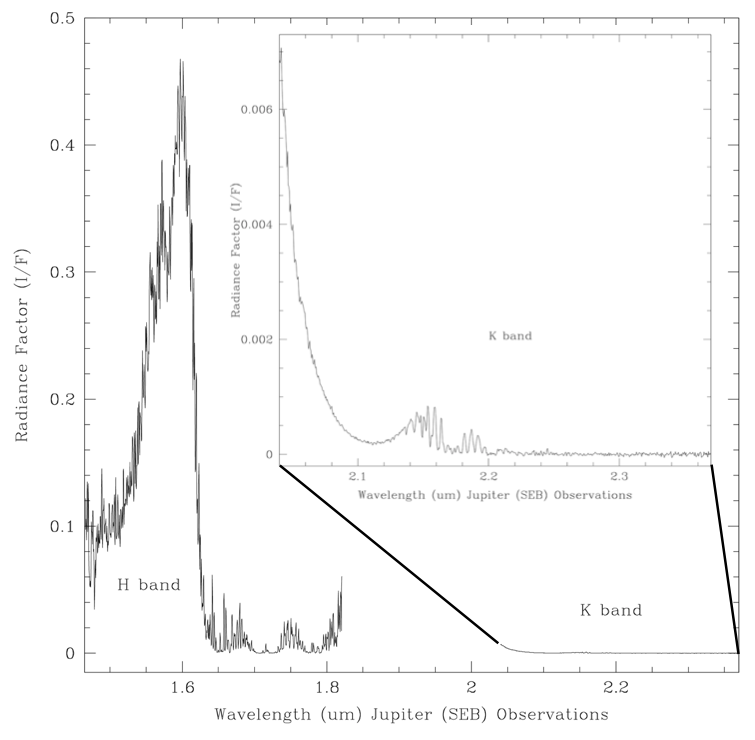}
\caption{The observed H and K bands spectra of Jupiter\textquotesingle s South Equatorial Belt.
}
\end{figure}

\thispagestyle{empty}

\begin{figure}[H]
\centering
\includegraphics[width=\columnwidth]{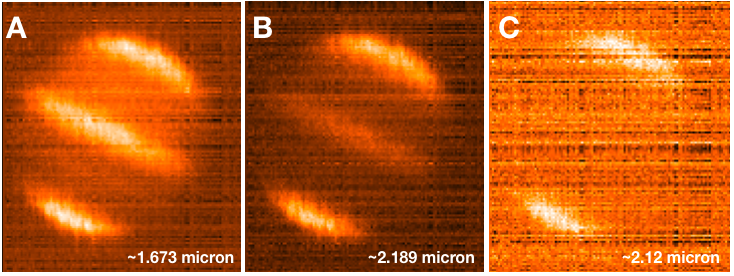}
\caption{Near-Infrared spectral 3-D cube images of Jupiter. They show different patterns of methane absorption in H band (A) and K band (B, C).
}
\end{figure}

\section{VSTAR-ATMOF model settings}
\label{S:3}

Versatile Software for Transfer of Atmospheric Radiation (VSTAR) modular package [12] and ATMOspheric Fitting (ATMOF) routine code [13] was used for our modelling. Jupiter\textquotesingle s atmosphere was divided into 46 layers of increasing altitude with specific pressure and temperature (P-T) profile. The chemical composition in each layer was expressed in terms of gas mixing ratios of molecules in the atmosphere. Ammonia and methane are the main molecular absorbers in Jupiter\textquotesingle s atmosphere. Their mixing ratios in our models were chosen and adjusted to match the recent findings from Juno measurements [14] and in situ measurements of the Galileo Probe Mass Spectrometer [15, 16] (Figure 5). The P-T profile of Jupiter\textquotesingle s atmosphere is based on Voyager\textquotesingle s radio occultation experiments [1, 17] (Black curves in figures 6, 7). Voyager\textquotesingle s occultation based data in our models provided a broader view of the planet\textquotesingle s P-T profile around the equator, while the Galileo Probe descended into a single point (a hot spot) in Jupiter\textquotesingle s North Equatorial Belt (NEB) [18] and cannot be the representative of a broader regional value (Kedziora-Chudczer and Bailey [9]).

\thispagestyle{empty}

Dominant molecules of Jupiter\textquotesingle s atmosphere are molecular hydrogen and helium which may be the source of Rayleigh Scattering and also important source of collisional-induced-absorption in the planet\textquotesingle s dense gaseous atmosphere. This collision-induced absorption (CIA) gives rise to the spectrum of overlapping lines with an impression of a smooth absorption trough centred at around 2.08 $\mu$m. Since the atmosphere of Jupiter has about 89 $\%$ and 10 $\%$ hydrogen and helium "respectively", the collisional interactions between both, H$_{2}$-H$_{2}$ and H$_{2}$-He molecules need to be included in our models. It worth mentioning that CIA mostly affects the K band region as its effects in J and H bands are small due to low CIA absorption coefficients between two prominent bands. 

We have also assumed an equilibrium ortho/para H$_{2}$ ratio and H/He ratio of 0.898/0.102 in our models. The latest spectral line lists for ammonia and methane (HITRAN2016, Gordon et al. [19]) was used to compute our models. The far wing line shape for methane is modified and based upon our best fits tested for Jupiter (Table 2) (Hartmann et al. [20]). The clouds that influence spectra are formed from condensation of ammonia. We characterise these clouds by assuming optical properties of ammonia in terms of the refractive index of the cloud particles set constant for all wavelengths to be n=1.43 and k=0.02i for the real (n) and imaginary (k) parts "respectively" as extracted from the literature [9].

\begin{figure}[H]
\centering
\includegraphics[width=\columnwidth]{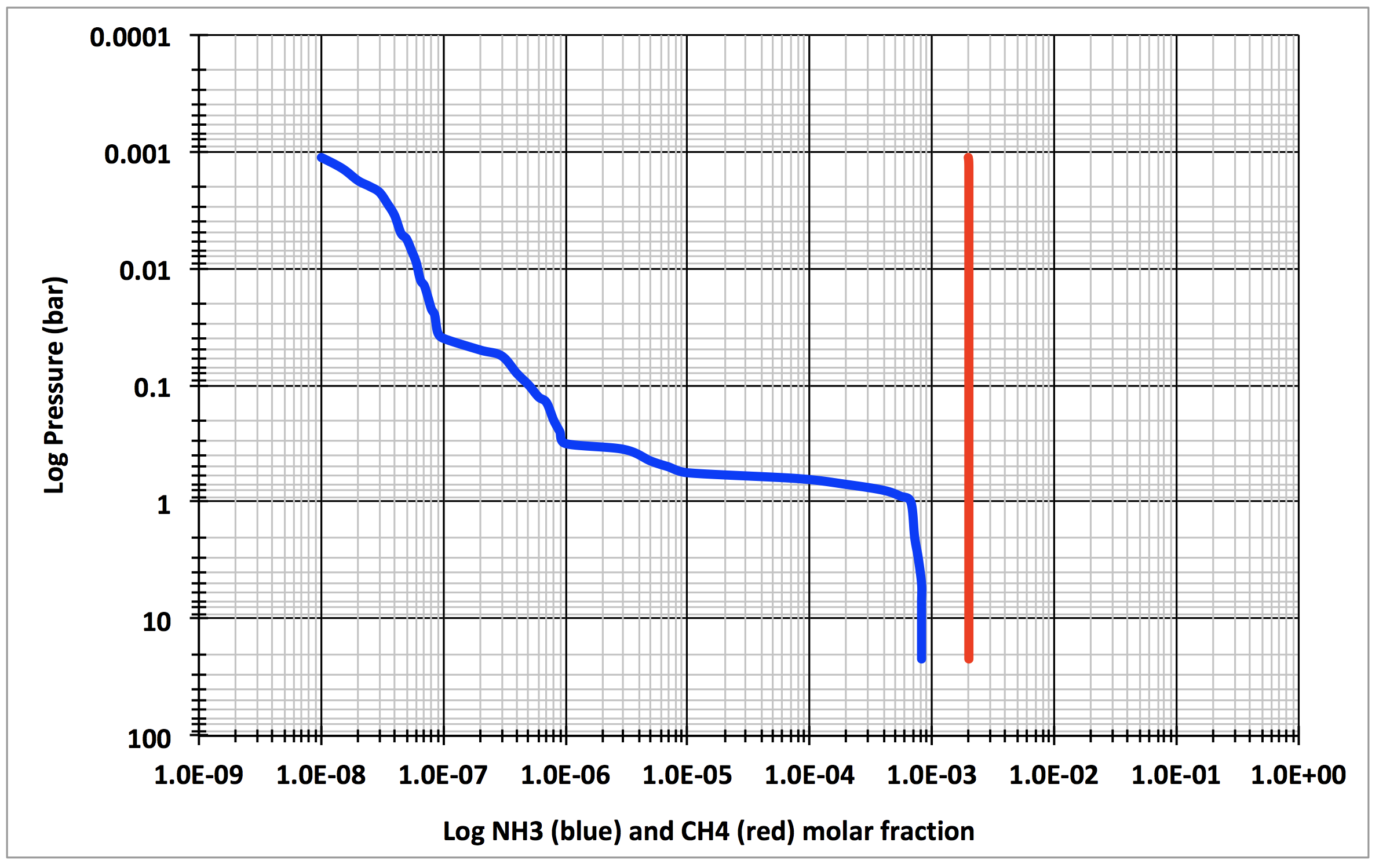}
\caption{Ammonia (blue) and methane (red) mixing ratio profiles as a function of pressure used in our models. 
}
\end{figure}

In our models the initial parameters are assigned to the distribution of opacities in the three thin clouds at different pressure levels using previously published results [9]. We then performed fitting for two cloud parameters, base pressures and opacity, by running atmospheric models using VSTAR, and performing least square fitting with ATMOF. In total, six parameters were modelled in ATMOF, the base pressure and opacity for each of three clouds, including upper cloud/haze, mid-cloud and lower-cloud as shown in Table 2. 

\thispagestyle{empty}

The final fitted values have been calculated for cloud base pressures and opacities. We did not fit for other parameters such as line shape parameters ($\sigma_{1}$, $\sigma_{2}$ , $A_{1}$), particle size (x) and cloud effective variables (distribution width, $\sigma$) for the sake of computational efficiency. Our cloud models along with P-T profile shown in Figures 6 and 7, simplified as a combined two cloud model layer with a single cloud deck pressure and opacity. In other words, the opacity of each cloud deck in our models divided between 2 layers. In reality each cloud opacity (the red and blue horizontal lines) shown in figures 6 and 7 are a combination of 2 cloud layers represented as one. These clouds may be in fact more extended vertically than represented as the reflectivity and scattering depend on the opacity of the cloud tops.

\begin{figure}[H]
\centering
\includegraphics[width=\columnwidth]{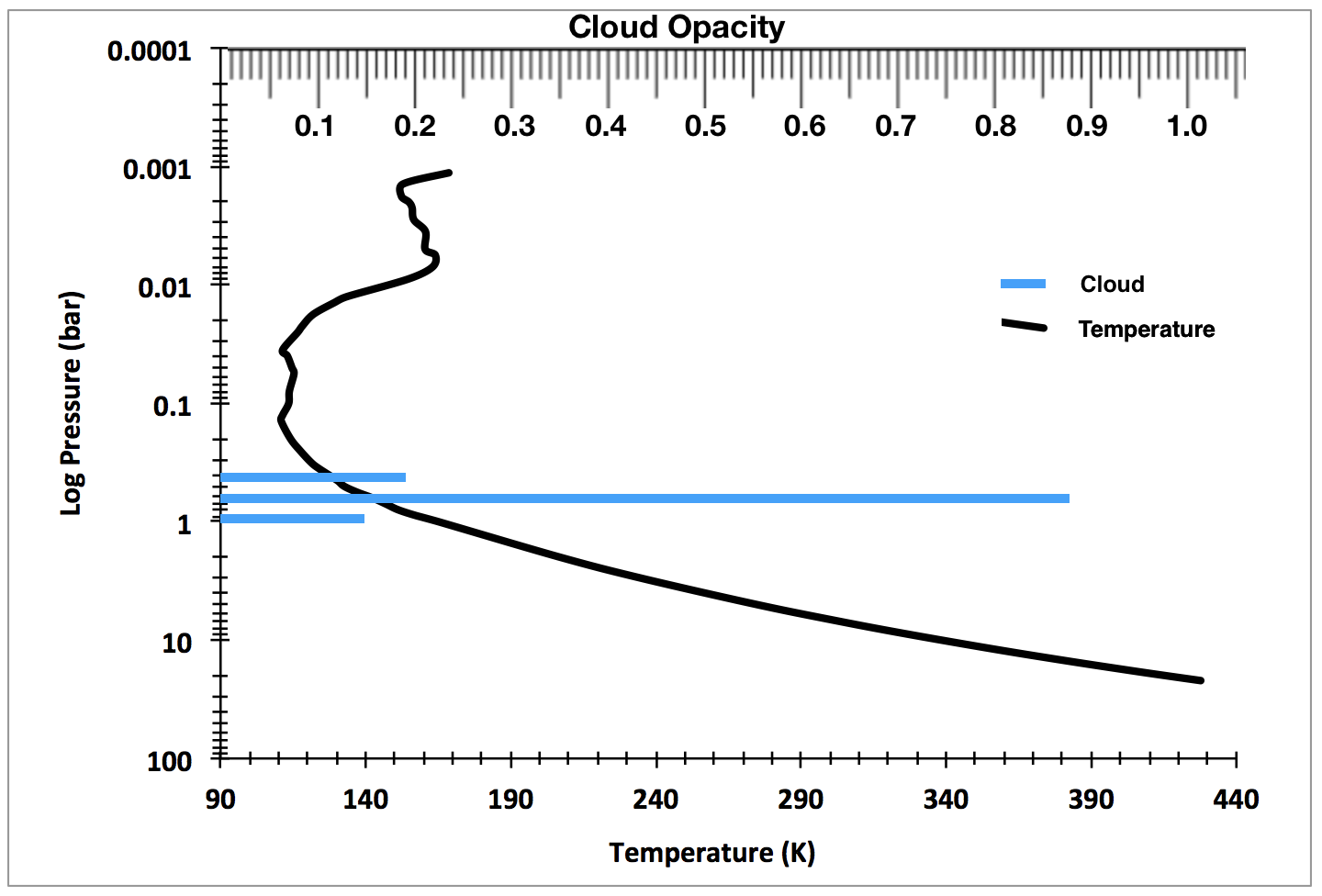}
\caption{Jupiter\textquotesingle s P-T profiles (black), best fitting cloud heights (base pressure of the cloud) and optical depths (blue) for H band. The resulting base pressures for the lower cloud, mid-cloud and the top haze are 0.94, 0.677 and 0.42 bar "respectively".
}
\end{figure}

\thispagestyle{empty}

\begin{figure}[H]
\centering
\includegraphics[width=\columnwidth]{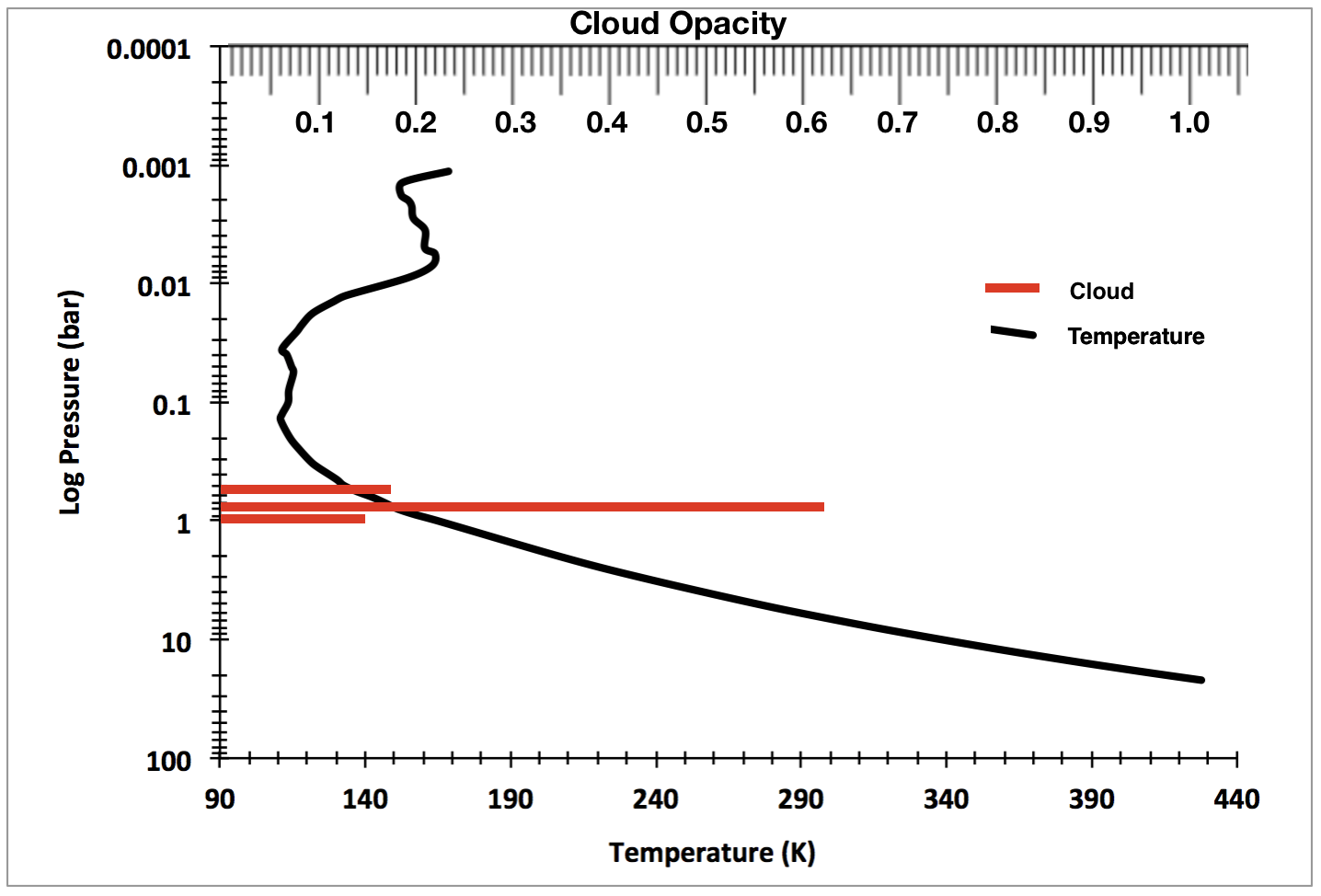}
\caption{Jupiter\textquotesingle s P-T profiles (black), best fitting cloud heights (base pressure of the cloud) and optical depths (red) for K band. The resulting base pressure for the lower cloud, mid-cloud and the top haze are 0.94, 0.79 and 0.542 bar "respectively".
}
\end{figure}

\thispagestyle{empty}

\section{Results and discussion}
\label{S:4}

The best fit spectral models for H and K bands along with their residuals are shown in Figure 8 (H band) and Figure 9 (K band). The best fitting parameters agree within their error bars between both bands, except for the pressure of the upper haze which indicates 0.122 bar difference. This is justifiable as the pressure variation in K band region is mostly affected by gaseous opacity of hydrogen rather than methane while in the H band region it is affected mostly by gaseous methane opacity. The best fit for H band shown in Figure 8 did not fit well in the peak window regions between 1.596 and 1.618 $\mu$m. This could be due to 1) incomplete methane line lists and variations in their transitional properties in the deeper (warmer) regions of Jupiter\textquotesingle s troposphere. 2) Strong absorption of the solar photons on top of the extended clouds as result of weak reflected signal of particles at allocated pressures (P $<$ 0.7 bar). 3) Changes in line shape profiles, other than we assumed in our models, capable of changing the spectral shape in both the centre and wings and 4) changes in methane gaseous mixing ratio profile in Jupiter\textquotesingle s atmosphere. We are confident to say that the first three scenarios are more probable. This is because we did not see any significant changes as result of modifying the gaseous methane mixing ratio in our models. 

Our H band\textquotesingle s final spectral fits show an ammonia cloud based at 0.677 bar with optical thickness of 0.877. Our model also shows the upper thin cloud/haze based at 0.42 bar with the opacity of 0.192 (Figure 6; Table 2). These results are compatible with the findings of Bjoraker et al. [21, 22] about Jupiter\textquotesingle s SEB, a region with lower albedo than the neighbouring zones, thinner clouds (ammonia and ammonium hydrosulfide) and consequently lower reflectance, due to probably multiple scattering and more absorbent cloud. Our models also place the lower cloud at around 1 bar with the optical depth of 0.15.

\begin{table}
\caption{Cloud/haze, particle size and line shape parameters for the modelled H and K bands
}
\centering
\resizebox{\textwidth}{!}{
\tabcolsep 1.4 pt
{\renewcommand{\arraystretch}{1.2} 
\begin{tabular}{|l|ll|ll|ll|lll|c|}
\hline
\hline
Jupiter\textquotesingle s band   & 
\multicolumn{2}{c|}{Upper cloud/haze} &
\multicolumn{2}{c|}{Mid-cloud (ammonia)} &
\multicolumn{2}{c|}{Lower cloud (ammonium hydrosulfide)} &
\multicolumn{3}{c|}{Assumed Line Shape} \\
Region &
\multicolumn{2}{c|}{x=0.3 $\mu$m and $\sigma$=0.1 $\mu$m} &
\multicolumn{2}{c|}{x=1.3 $\mu$m and $\sigma$=0.5 $\mu$m} &
\multicolumn{2}{c|}{x=1.5 $\mu$m and $\sigma$=0.5 $\mu$m} &
\multicolumn{3}{c|}{} \\
\hline
 &
\multicolumn{1}{c|}{P (bar)} & \multicolumn{1}{c|}{$\tau$} & \multicolumn{1}{c|}{P (bar)} & \multicolumn{1}{c|}{$\tau$} & \multicolumn{1}{c|}{P (bar)} & \multicolumn{1}{c|}{$\tau$} &
\multicolumn{1}{c|}{$\sigma_{1}$} & \multicolumn{1}{c|}{$\sigma_{2}$} & \multicolumn{1}{c|}{A$_{1}$} \\
\hline
H band initial fit &
\multicolumn{1}{c|}{0.5} & \multicolumn{1}{c|}{0.153} & \multicolumn{1}{c|}{0.68} & \multicolumn{1}{c|}{0.845} & \multicolumn{1}{c|}{2.5} & \multicolumn{1}{c|}{0.9} &
\multicolumn{1}{c|}{100} & \multicolumn{1}{c|}{200} & \multicolumn{1}{c|}{1.0} \\
Final fit &
\multicolumn{1}{c|}{0.42 $\pm$ 0.05} & \multicolumn{1}{c|}{0.192 $\pm$ 0.023} & \multicolumn{1}{c|}{0.677 $\pm$ 0.068} & \multicolumn{1}{c|}{0.877 $\pm$ 0.084} & \multicolumn{1}{c|}{0.94 $\pm$ 0.25} & \multicolumn{1}{c|}{0.15 $\pm$ 0.09} &
\multicolumn{1}{c|}{} & \multicolumn{1}{c|}{} & \multicolumn{1}{c|}{} \\
\hline
K band initial fit &
\multicolumn{1}{c|}{0.5} & \multicolumn{1}{c|}{0.153} & \multicolumn{1}{c|}{0.68} & \multicolumn{1}{c|}{0.845} & \multicolumn{1}{c|}{2.5} & \multicolumn{1}{c|}{0.9} &
\multicolumn{1}{c|}{100} & \multicolumn{1}{c|}{200} & \multicolumn{1}{c|}{1.0} \\
Final fit &
\multicolumn{1}{c|}{0.542 $\pm$ 0.050} & \multicolumn{1}{c|}{0.176 $\pm$ 0.015} & \multicolumn{1}{c|}{0.79 $\pm$ 0.23} & \multicolumn{1}{c|}{0.621 $\pm$ 0.295} & \multicolumn{1}{c|}{0.94 $\pm$ 0.25} & \multicolumn{1}{c|}{0.15 $\pm$ 0.09} &
\multicolumn{1}{c|}{} & \multicolumn{1}{c|}{} & \multicolumn{1}{c|}{} \\
\hline
\hline

\end{tabular}
}
}

\label{tab:LR_results}
\end{table}



Our K band final spectral fits show an ammonia cloud based at 0.79 bar with the optical thickness of 0.621. Our model also show the haze based at 0.542 bar with the opacity of 0.176 (Figure 7; Table 2). The best fit for K band as shown in Figure 9, fits the data well except in the regions around 2.08, 2.16 to 2.19 $\mu$m. The 2.08 $\mu$m region is dominated by collision-induced-absorption while 2.16-2.19 $\mu$m region is dominated by methane absorption. Decreased reflectivity in our model (around 2.08 $\mu$m) along with shortness in reaching the extremities (~2.16 to 2.19 $\mu$m) may be related to assumed parameters such as line shape, particle size and cloud effective variables, which may improve our models if fitted. Other possible scenarios to consider are the incomplete (broadened) methane absorption lines and different P-T profile at the upper atmosphere as a cause of more stretched atmospheric scale height in our models. Note, we do not discuss the fitting discrepancies observed between 2.04 and 2.07 $\mu$m in our K band model. This is because 1) regions close to the beginning and the end of each spectral band may have instrumental flux calibration issues and 2) investigating the hydrogen collision-induced opacity in higher pressures of Jupiter\textquotesingle s troposphere is out of the scope of this paper.

\thispagestyle{empty}
  
In general, both H and K band models show the same base pressure as well as optical depth for the observed lower cloud. The difference lies in the NH$_{3}$ cloud region with different base pressure heights and opacities. As already mentioned above, we did not fit the particle size parameter in our models for the sake of model complexity and computational efficiency and assumed clouds at different pressures with the single dominant size of particles with the set distribution. More realistic clouds may have more complex particle distribution. Therefore, considering 1) complex particle size distribution of the clouds in reality; 2) the 35 minute observing time difference between H and K bands, capable of monitoring different location on the planet as result of its fast rotation period (~9.8 hours); and 3) weather related vertical wind decay profile variations with depth [23, 24, 25, 26] indicative of cold air under the cyclonic belts [27], we may conclude that the westward retrograde atmospheric jet-stream along with strong vertical wind shear in higher altitudes can affect the higher parts of the vertically extended clouds more than the lower parts. This in turn will result in thorough sorting of particles in higher cloud altitudes than the lower. These vigorous fast moving and circulating currents in higher parts of the vertically extended clouds drag small pockets of larger particles, from more stratified cloud deck, to the lower pressures (higher altitudes) as result of Kelvin-Helmholtz instabilities. 

Methane is well mixed in Jupiter\textquotesingle s atmosphere (up to methane homopause), but there are zonal (latitudinal) differences in the mixing ratio of ammonia which is sensitive to the pressure ranges higher than $\sim$700 mbar as pointed by Bolton et al. [14] and Ingersoll et al. [28]. As shown in Figure 8, a discrepancy of more absorption is present in the modelled spectrum in H band region of ammonia absorption centred around 1.56 $\mu$m. This inconsistency may be interpreted as the lack of pressure sensitivity in the ammonia mixing ratio above the NH$_{3}$ cloud tops (P $<$ 0.7 bar) in our modelled spectral regions as discussed in Ingersoll et al. [28]. Our H band model shows the ammonia cloud deck at 0.677 bar, which is within one standard deviation of the minimum sensitivity pressure range (0.7). Therefore, the higher absorption in our model is not related to the lowest sensitivity pressure range. However, we blame the complex particle size distribution (as consequences of masked ammonia cloud particles in allocated pressure range) and their complicated refractive index profile. 

\thispagestyle{empty}

\begin{figure}[H]
\centering
\includegraphics[width=\columnwidth]{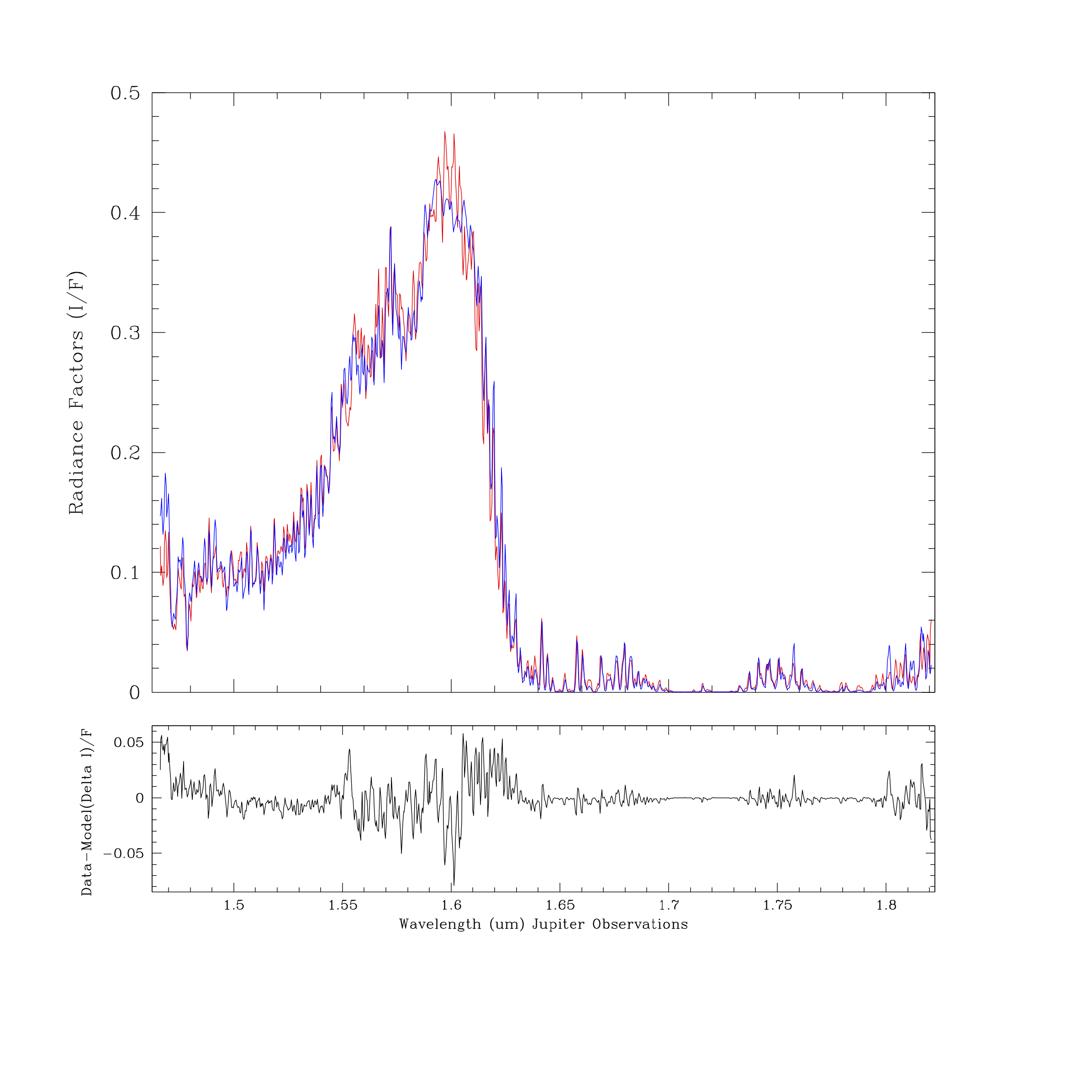}
\caption{Top box shows the observed spectrum (red) along with fitted model spectrum (blue) for the H band in Jupiter\textquotesingle s SEB. The bottom box shows the residuals (modelled radiance factor (I/F) subtracted from the observed). 
}
\end{figure}

\begin{figure}[H]
\centering
\includegraphics[width=\columnwidth]{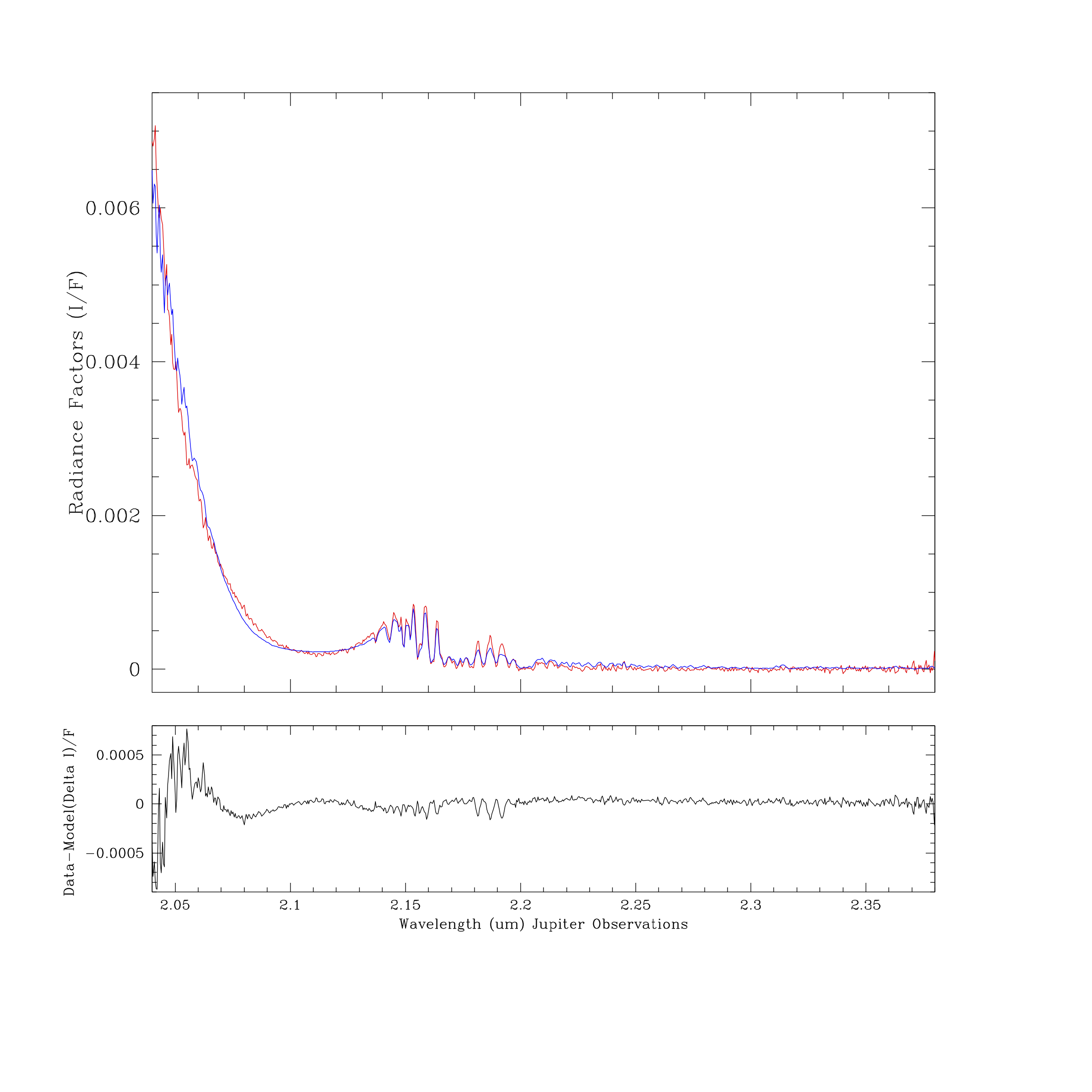}
\caption{ Top box shows the observed spectrum (red) along with fitted model spectrum (blue) for K band in Jupiter\textquotesingle s SEB. The bottom box shows the residuals (modelled radiance factor (I/F) subtracted from the observed).
}
\end{figure}

\section{Conclusions}
\label{S:5}

The models illustrated in this paper provide good fitting values to the clouds in H band of Jupiter\textquotesingle s SEB except in the peak window region between 1.596 and 1.618 $\mu$m. We interpreted the discrepancies as incomplete methane line lists and variations in their transitional properties in warmer deeper pressures, along with strong absorption of the solar photons on top of the extended clouds at allocated pressures (P $<$ 0.7 bar). We also hold responsible any sort of variations in the line shape profiles, capable of changing the spectral shape in both centre and neighbouring absorption wings.

\thispagestyle{empty}

The poor fitting of more ammonia absorption also presented in the H band centred around 1.56 $\mu$m. This is related to the lack of pressure sensitivity in the ammonia mixing ratio above the NH$_{3}$ cloud tops (P $<$ 0.7 bar). The higher absorption is related to the complex particle size distribution as result of masked ammonia cloud particles along with their complicated refractive index profile.

Our K band model fits the data well except in the regions around 2.08, 2.16 to 2.19 $\mu$m. Decreased reflectivity in our model (around 2.08, ~2.16 to 2.19 $\mu$m) may be related to unaccounted parameters such as line shape, particle size and cloud effective variables, which may improve our models if fitted.

Resulted variations in the cloud pressure heights and their opacity suggest that the modelled physical conditions correctly produce the majority of the observed H and K band spectrum in Jupiter\textquotesingle s SEB, however small residual discrepancies are evident at certain wavelengths, and is related to the presence of vertically extended ammonia clouds affecting the observed reflectivity (scattering) and opacity. Good agreement of cloud parameters within the fit errors obtained for both spectral bands, suggest the same consistent system of clouds in both spectral regions.

Modelling of the other latitudinal and longitudinal regions of Jupiter in three spectral bands of J, H and K is in progress to cover the global mapping of the clouds/haze across the planetary disk for better understanding of Jupiter\textquotesingle s atmospheric characteristics and features. 

\thispagestyle{empty}

\section*{Acknowledgements}

B. Karamiqucham acknowledges financial support of the Commonwealth of Australia for the project as Australian Government\textquotesingle s Research Training Program (RTP). We acknowledge the support of the Anglo-Australian Observatory in scheduling the observation run presented in this paper using IRIS2 service observing instruments.

\section{References}
\label{S:7}

1. Lindal, G. F., Wood, G. E., Levy, G. S., Anderson, J. D., Sweetnam, D. N., Hotz, H. B., Buckles, B. J., Holmes, D. P., Doms, P. E., Eshleman, V. R. Et al., “The atmosphere of Jupiter: an analysis of the Voyager radio occultation measurements”, \textit{Journal of Geophysical Research}, Vol. 86, 1981, pp.8721-8727.\\

\noindent2.  Atreya, S. K., Wong, M. H., Owen, T. C., Mahaffy, P. R., Niemann, H. B., de Pater, I., Drossart, P., and Encrenaz, Th., “A comparison of the atmospheres of Jupiter and Saturn: Deep atmospheric composition, cloud structure, vertical mixing and origin”, \textit{Planetary and Space Science}, Vol. 47, 1999, pp.1243-1262.\\

\noindent3.  Atreya, S.K., A.S. Wong, K.H. Baines, M.H. Wong, and T.C. Owen, \lq\lq Jupiter\textquotesingle s ammonia clouds: Localized or ubiquitous?,\rq\rq \textit{Planetary and Space Science}, Vol. 53, 2005, pp.498-507.\\

\noindent4. de Pater, I., D. Dunn, P.N. Romani, and K. Zahnle, “Reconciling Galileo probe data and ground-based radio observations of ammonia on Jupiter”, \textit{Icarus}, Vol. 149, 2001, pp.66-78.\\

\noindent5. Irwin, P.G.J., “Cloud structure and composition of Jupiter's atmosphere”, \textit{Surveys in Geophysics}, Vol. 20, 1999, pp.505-535.\\

\noindent6. Irwin, P.G.J. and U. Dyudina, “The retrieval of cloud structure maps in the equatorial region of Jupiter using a principal component analysis of Galileo/NIMS data”, \textit{Icarus}, Vol. 156, 2002, pp.52-63.\\

\noindent7. Irwin, P. G. J., Weir, A. L., Smith, S. E., Taylor, F. W., Lambert, A. L., Calcutt, S. B., Cameron- Smith, P. J., Carlson, R. W., Baines, K., Orton G. S. Et al., “Cloud structure and atmospheric composition of Jupiter retrieved from Galileo near-infrared mapping spectrometer real-time spectra”, \textit{Journal of Geophysical Research}, Vol. 103, 1998, pp.23001-23021.\\

\noindent8. Irwin, P.G.J., A.L. Weir, F.W. Taylor, S.B. Calcutt, and R.W. Carlson, “The origin of belt/ zone contrasts in the atmosphere of Jupiter and their correlation with 5-$\mu$m opacity”, \textit{Icarus}, Vol. 149, 2001, pp.397-415.\\

\thispagestyle{empty}

\noindent9. Kedziora-Chudczed, L. And Bailey, J., “Modelling the near-infrared spectra of Jupiter using line-by-line methods”, \textit{Monthly Notices of Royal Astronomical Society}, Vol. 414, 2011, pp.1483-1492.\\

\noindent10. Tinney, C. G., Ryder, S. D., Ellis, S. C., Churilov, V., Dawson, J., Smith, G. A., Waller,  L., Whittard, J. D et al., “IRIS2: a working infrared multi-object spectrograph and camera”, Proc. SPIE 5492, \textit{Ground-based Instrumentation for Astronomy}, Vol. 5492, 2004, pp.998-1009.\\

\noindent11. Shortridge, K., Meatheringham, S. J., Carter, B. D. And Ashley, C. B., “Making the Figaro data reduction system portable”, \textit{Publication of the Astronomical Society of Australia}, Vol. 12, 1995, pp.244-247.\\

\noindent12. Bailey, J. and Kedziora-Chudczer, L., “Modelling the spectra of planets, brown dwarfs and stars using VSTAR”, \textit{Monthly Notices of Royal Astronomical Society}, Vol. 419, 2012, pp.1913-1929.\\

\noindent13. Cotton, D. V., Bailey, J. and Kedziora-Chudczer, L., “Atmospheric modelling for the removal of telluric features from infrared planetary spectra”, \textit{Monthly Notices of Royal Astronomical Society}, Vol. 439, 2014, pp.387-399.\\

\noindent14. Bolton, S. J., Adriani, A., Adumitroaie, V., Allison, M., Anderson, J., Atreya, S., Bloxham, J. et al., “Jupiter’s interior and deep atmosphere: The initial pole-to-pole passes with the Juno spacecraft”, \textit{Science}, Vol. 356, 2017, pp.821-825.\\

\noindent15. Wong, M. H., Mahaffy, P. R., Atreya, S. K., Niemann, H. B. and Owen, T. C., “Updated Galileo probe mass spectrometer measurements of carbon, oxygen, nitrogen and sulfur on Jupiter”, \textit{Icarus}, Vol. 171, 2004, pp.153-170.\\

\noindent16. Moses, J. I., T. Fouchet, B. Bezard, G. R. Gladstone, E. Lellouch, and H. Feuchtgruber (2005), “Photochemistry and diffusion in Jupiter’s stratosphere: Constraints from ISO observations and comparisons with other giant planets” \textit{Journal of Geophysical Research}, Vol. 110, 2005, E08001.\\

\noindent17. Lindal, G. F., “The atmosphere of Neptune: An analysis of radio occultation data acquired with Voyager 2”, \textit{Astronomical Journal}, Vol. 103, 1992, pp.967-982.\\

\noindent18. Seiff, A., Kirk, D. B., Knight, T. C. D., Young, R. E., Mihalov, J. D., Young, L. A., Milos, F. S., Schubert, G., Blanchard, R. C. and Atkinson, D., “Thermal structure of Jupiter’s atmosphere near the edge of a 5-$\mu$m hot spot in the north equatorial belt”, \textit{Journal of Geophysical Research}, Vol. 103, 1998, pp.22857-22889.\\

\noindent19. Gordon, I. E., Rothman, L. S., Hill, C., Kochanov, R. V., Tan, Y., Bernath, P. F. et al.,  “The HITRAN2016 molecular spectroscopic database”, \textit{Journal of Quantitative Spectroscopy and Radiative Transfer}, Vol. 203, 2017, pp.3-69.\\

\thispagestyle{empty}

\noindent20. Hartmanna, J. M., Bouleta, C., Brodbecka, C., van Thanha, N., Fouchetb, T. and Drossartb, P., “A far wing lineshape for H2 broadened CH$_{4}$ infrared transitions”, \textit{Journal of Quantitative Spectroscopy and Radiative Transfer}, Vol. 72, 2002, pp.117–122.\\

\noindent21. Bjoraker, G. L., Wong, M. H., de Pater, I. and Ádámkovics, M. “Jupiter’s deep cloud structure revealed using Keck observations of spectrally resolved line shapes”, \textit{The Astrophysical Journal}, Vol. 122, 2015, pp.810-819.\\

\noindent22. Bjoraker, G. L., Wong, M. H., de Pater, I., Hewagama, T., Ádámkovics, M. And Orton, G. S. “The gas composition and deep cloud structure of Jupiter’s Great Red Spot”, \textit{The Astrophysical Journal}, Vol. 156, 2018, pp.101-115.\\

\noindent23. Ingersoll, A. P. and Cuzzi, J. N., “Dynamics of Jupiter’s cloud bands”, \textit{Journal of Atmospheric Sciences}, Vol. 26, 1969, pp.981-985.\\

\noindent24. Barcilon, A. and Gierasch, P. J., “A moist, Hadley cell model for Jupiter’s cloud bands”, \textit{Journal of Atmospheric Sciences}, Vol. 27, No. 4, 1970, pp.550-560.\\

\noindent25. Guillot, T., Miguel, Y., Militzer, B., Hummard, W. B., Kaspi, Y., Galanti, E., Cao, H. et al., “A suppression of differential rotation in Jupiter’s deep interior”, \textit{Nature}, Vol. 555, 2018, pp.227-230.\\

\noindent26. Kaspi, Y., Galanti, E., Hubbard, W. E., Stevenson, D. J., Bolton, S. J., Iess, L., Guillot, T. et al., “Jupiter’s atmospheric jet streams extend thousands of kilometers deep”, \textit{Nature}, Vol. 555, 2018, pp.223-226.\\

\noindent27. Gierasch, P. J., Conrath, B. J. and Magalhaes, J. A., “Zonal mean properties of Jupiter upper troposphere from Voyager infrared observations”, \textit{Icarus}, Vol. 67, No. 3, 1986, pp.456-483.\\

\noindent28. Ingersoll, A. P., Adumitroaie, V., Allison, M. D., Atreya, S. et al., “Implications of the ammonia distribution on Jupiter from 1 to 100 bars as measured by Juno microwave radiometer”, \textit{Geophysical Research Letters}, Vol. 44, 2017, pp.7676-7685.\\

\thispagestyle{empty}













\end{document}